\documentclass[aps,pra]{revtex4}

\usepackage{amssymb}
\usepackage{amsmath}
\usepackage[xdvi,dvips]{graphicx}
\usepackage{graphicx}

\newcommand{\la}{\lambda}

\newcommand{\br}{{\bf r}}

\newcommand{\bu}{{\bf u}}

\begin{document}

\title{Interaction of oblique dark solitons in two-dimensional supersonic nonlinear Schr\"odinger flow}

\author{E. S. Annibale}
\email{annibale@if.usp.br}
\author{A. Gammal}
\email{gammal@if.usp.br}


\affiliation{
Instituto de F\'{\i}sica, Universidade de S\~{a}o Paulo,
05508-090, S\~{a}o Paulo, Brazil}




\date{\today}

\begin{abstract} 
We investigate the collision of two oblique dark solitons in the two-dimensional supersonic
nonlinear Schr\"odinger flow past two impenetrable obstacles. We numerically show that this collision 
is very similar to the dark solitons collision in the one dimensional case. 
We observe that it is practically elastic and we measure the shifts of the solitons positions
after their interaction.

\end{abstract}

\maketitle

1. Two-dimensional (2D) oblique dark solitons are unstable with respect to transverse 
perturbations \cite{kp-1970,zakharov-1975,kt-1988} and therefore their interaction with each 
other is not of much interest from practical point of view. However, it has been found
\cite{egk-06} that such solitons generated in the flow of an atomic Bose-Einstein condensate (BEC)
past an obstacle behave as effectively stable. Such a behavior was explained in \cite{kp-08}
as a result of the
transition from absolute instability of 2D solitons to their convective instability for large 
enough velocities of the flow in the reference frame attached to the obstacle, so that 
unstable modes are convected by the flow along the solitons from the region around the obstacle. 
The condition for convective instability of these dark solitons was derived in \cite{kk-11}.
This phenomenon has a general nature and its nonlinear optics counterpart has been discussed 
in \cite{kgegk-08}. Recently, experimental observations in Bose-Einstein condensate of exciton-polaritons have 
indeed demonstrated the existence of stable oblique dark solitons in a superfluid flow past an 
obstacle \cite{amo-2010}. Hence, interaction of such effectively stable oblique dark solitons 
becomes a question of considerable interest and it will be addressed in this paper.

\vspace{1cm}

2. Oblique dark solitons in a superfluid are described very well \cite{egk-06}
as stationary solutions of the defocusing nonlinear Schr\"odinger equation (NLS)
\begin{equation}\label{1-1}
    i\psi_t=-\frac12\Delta\psi+|\psi|^2\psi \,,
\end{equation}
which is written here in standard dimensionless units and 
$\Delta \equiv \partial^2_x+\partial^2_y $.
Its transformation to a ``hydrodynamic form'' by means of the substitution
\begin{equation}\label{1-2}
    \psi(\br,t)=\sqrt{n(\br,t)}\exp\left(i\int^{\br'}\bu(\br',t)d\br'\right)
\end{equation}
yields the system
\begin{equation}\label{1-3}
    n_t+\nabla\cdot(n\bu)=0,
\end{equation}
\begin{equation}\label{1-4}
    \bu_t+(\bu\cdot\nabla)\bu+\nabla n+\nabla\left(\frac{(\nabla n)^2}{8n^2}
    -\frac{\Delta n}{4n}\right)=0,
\end{equation}
where $n$ is the density of the fluid and $\bu$ denotes its velocity field.

In a stationary case $(n_t=0,\,\bu_t=0)$ this system takes the form
\begin{equation}\label{2-1}
\begin{split}
   (n u)_x+(n v)_y=0,\\
   uu_x+vu_y+n_x+\left(\frac{n_x^2+n_y^2}{8n^2}-
   \frac{n_{xx}+n_{yy}}{4n}\right)_x=0,\\
   uv_x+vv_y+n_y+\left(\frac{n_x^2+n_y^2}{8n^2}-
   \frac{n_{xx}+n_{yy}}{4n}\right)_y=0,
   \end{split}
\end{equation}
where we have introduced the components $\bu=(u,v)$ of the velocity field.
It should be solved with the boundary conditions
\begin{equation}\label{2-2}
   n=1,\quad u=M,\quad v=0\quad\text{at}\quad |x|\to\infty,
\end{equation}
which means that there is a uniform flow of a superfluid with constant velocity
$\bu=(M,0)$ at infinity. Since in our dimensionless units the sound velocity
at infinity is equal to unity, the incoming velocity coincides with the
Mach number $M$. The soliton solution of this problem was found in
\cite{egk-06} and it can be written as
\begin{equation}\label{2-3}
    n(x,y)=1-\frac{1-M^2\sin^2\theta}
    {\cosh^2[\sqrt{1-M^2\sin^2\theta}(x\sin\theta-y\cos\theta+y_0\cos\theta)]},
\end{equation}
\begin{equation}\label{2-4}
    u(x,y)=M\left[1+\sin^2\theta\left(\frac1{n(x,y)}-1\right)\right],\quad
    v(x,y)=-M\sin\theta\cos\theta\left(\frac1{n(x,y)}-1\right),
\end{equation}
\begin{figure}[t!!]
\begin{center}
\includegraphics[width=10cm]{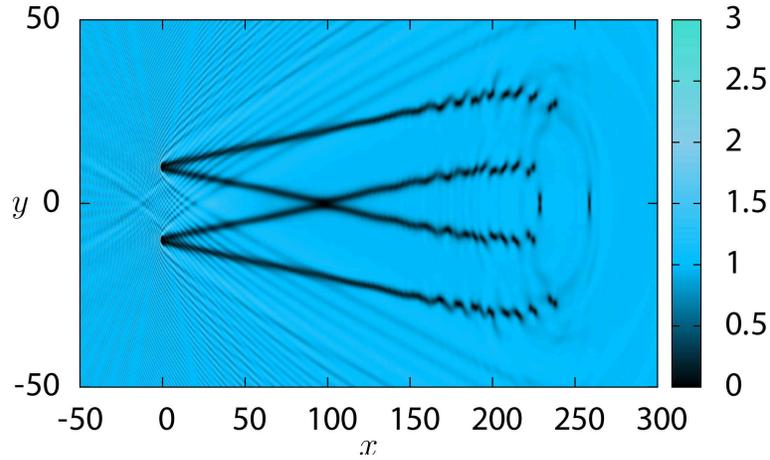}
\caption{\label{fig1} Interaction of two oblique dark solitons generated in the 
flow of a superfluid past two impenetrable obstacles. 
The flow is from the left to the right with $M=5$. One obstacle is
located at $(x_{10},y_{10})=(0,-10)$ and the other one at
$(x_{20},y_{20})=(0,10)$~.
Colors indicate the magnitude of the density.}
\end{center}
\end{figure}
where $\theta$ is the angle between the oblique soliton and the horizontal axis
and $y_0$ is its intersection point with the $y$ axis. 
The transformation (\ref{1-2}) implies that the flow is potential
(vorticity free) so that the velocity field {\bf u} can be represented as a
gradient of the phase
\begin{equation}\label{2-5}
    \phi(x,y)=Mx-\arctan\frac{M\sin\theta}{\sqrt{1-M^2\sin^2\theta}
    \tanh[\sqrt{1-M^2\sin^2\theta}(x\sin\theta-y\cos\theta+y_0\cos\theta)]}.
\end{equation}
Correspondingly, the wave function of the oblique soliton
reads
\begin{equation}\label{3-1}
    \psi(x,y)=\left\{\sqrt{1-M^2\sin^2\theta}
    \tanh\left[\sqrt{1-M^2\sin^2\theta}(x\sin\theta-y\cos\theta+y_0\cos\theta)\right]
    -iM\sin\theta\right\}\exp(iMx).
\end{equation}
This formula describes the oblique solitons generated by the flow of a superfluid
past an impenetrable obstacle. 
It is clear from this formula that such solitons can only be generated inside 
the Mach cone,
\begin{equation}\label{3-2}
    -\arcsin(1/M)<\theta<\arcsin(1/M).
\end{equation}

\vspace{1cm}

3. If there are several obstacles in the flow of a superfluid, then several dispersive
shocks are generated which decay far enough from the obstacles into oblique dark
solitons. When such space solitons overlap, they interact with each other
and their behavior in the overlap region is of considerable interest.
We have simulated the interaction of oblique solitons numerically and the results are
shown in Fig.~1.
As we see, two pairs of dark solitons are generated, two of these solitons interact with each
other in the region far enough from the obstacles and the end points of the solitons
decay into vortices. It is remarkable that the interaction is practically
elastic---no new solitons or radiation are visible. The only visible result of the
interaction is a shift of the solitons positions after their interaction. This behavior
is typical for the systems described by so-called completely integrable evolution
equations (see, e.g. \cite{zmnp-80}). Although there is nothing known about
complete integrability of the system (\ref{2-1}), it has well-known limiting
cases when it reduces to completely integrable equations (see, e.g.,
\cite{egk-06,ek-06,egk-07}):
first, the limit of shallow solitons, when the system reduces to the Korteweg-de Vries
(KdV) equation, and, second, the hypersonic limit $M\gg1$ when it reduces to 1D
NLS equation. This indicates that the system (\ref{2-1}) is in a sense ``close''
to the completely integrable equations and therefore it demonstrates similar
behavior. 
In this work, we concentrate only on the study of 
deep solitons so we shall consider the hypersonic limit and derive formulae for the
corresponding shifts of the solitons position. 

\vspace{1cm}

\begin{figure}[t]
\begin{center}
\includegraphics[width=6cm,height=6cm,clip]{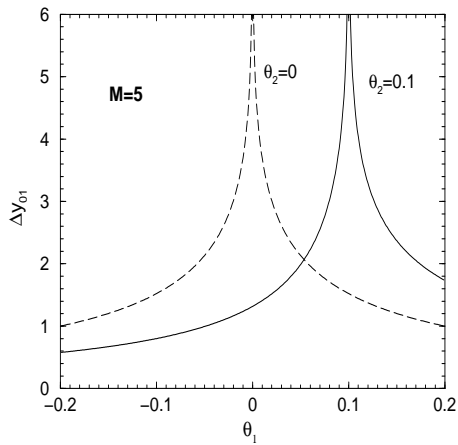}
\caption{Shifts of the oblique solitons positions as functions of
their slope angle. The second soliton has the slope angles:
$\theta_{2}=0$ (dashed line); $\theta_{2}=0.1$ (solid line).
The Mach number is equal to $M=5$.}
\end{center}\label{fig2}
\end{figure}
4. Let us consider the hypersonic limit $M\gg1$. As was shown in \cite{ek-06,ekkag-09},
in the leading order of the expansion with respect to the small parameter
$1/M\ll1$ the system (\ref{2-1}) reduces to
\begin{equation}\label{4-1}
    \begin{split}
    n_T+(n v)_Y=0,\\
    v_T+vv_Y+n_Y+\left(\frac{(n_Y)^2}{8n^2}-\frac{n_{YY}}{4n}\right)_Y=0,
    \end{split}
\end{equation}
and
\begin{equation}\label{4-2}
    u_{1,T}+vu_{1,Y}=0,
\end{equation}
where we have introduced the notation
\begin{equation}\label{4-3}
    u=M+u_1+O(1/M),\quad T=\frac{x}M,\quad Y=y.
\end{equation}
The system (\ref{4-1}) is nothing but the hydrodynamic form of the
1D NLS equation
\begin{equation}\label{4-4}
    i\Psi_T+\tfrac12\Psi_{YY}-|\Psi|^2\Psi=0
\end{equation}
for the variable
\begin{equation}\label{4-5}
    \Psi(Y,T)=\sqrt{n(Y,T)}\exp\left(i\int^Y v(Y',T)dY'\right).
\end{equation}
As it is well known, the NLS equation (\ref{4-4}) is completely integrable,
it has exact multi-soliton solutions and interaction of two solitons
was already studied in the classical paper  \cite{zs-1973}.
The single soliton solution of the equation (\ref{4-4}) is parameterized
conveniently by the value $\la$ of the associated Zakharov-Shabat
spectral problem and after returning to the $x,\,y$ coordinates
it takes the form
\begin{equation}\label{4-6}
    n(x,y)=1-\frac{1-\la^2}{\cosh^2[\sqrt{1-\la^2}(y-\la x/M-y_0)]}
\end{equation}
so that
\begin{equation}\label{4-7}
    \la\cong\theta M\quad \hbox{for}\, |\theta|\ll 1.
\end{equation}
It is natural that Eq.~(\ref{2-3}) reduces to (\ref{4-6}) in the limit
(\ref{4-7}).

If there are two oblique solitons in the superfluid, then they are
characterized by two parameters $\la_{1,2}$ corresponding
to different angles $\theta_{1,2}\cong\la_{1,2}/M$ and by two different
``initial'' coordinates $y_{10},\,y_{20}$. We suppose that $\la_1>\la_2$
and $y_{10}<y_{20}$. Then the shifts $\Delta y_{i}$ of the asymptotic
``positions'' of the oblique solitons are described by the formulae
\cite{zs-1973,zmnp-80}
\begin{equation}\label{5-1}
    \Delta y_{1}=\frac1{2\nu_1}\ln\frac{(\la_1-\la_2)^2+(\nu_1+\nu_2)^2}
    {(\la_1-\la_2)^2+(\nu_1-\nu_2)^2},\quad
    \Delta y_{2}=-\frac1{2\nu_2}\ln\frac{(\la_1-\la_2)^2+(\nu_1+\nu_2)^2}
    {(\la_1-\la_2)^2+(\nu_1-\nu_2)^2},
\end{equation}
where
\begin{equation}\label{5-2}
    \la_i=M\theta_i,\quad \nu_i=\sqrt{1-\la_i^2},\quad i=1,2. 
\end{equation} These formulae describe the shifts for the case $M\gg1$.
 
The dependence of the shifts $\Delta y_{1}$ on $\theta$ [see Eqs. (\ref{5-1}),(\ref{5-2})]
for some values of the slope angle of the second soliton and $M=5$ is shown in Fig.~2.

\vspace{1cm}

\begin{figure}[t!!]
\begin{minipage}[b]{0.45\linewidth}
\includegraphics[width=\linewidth]{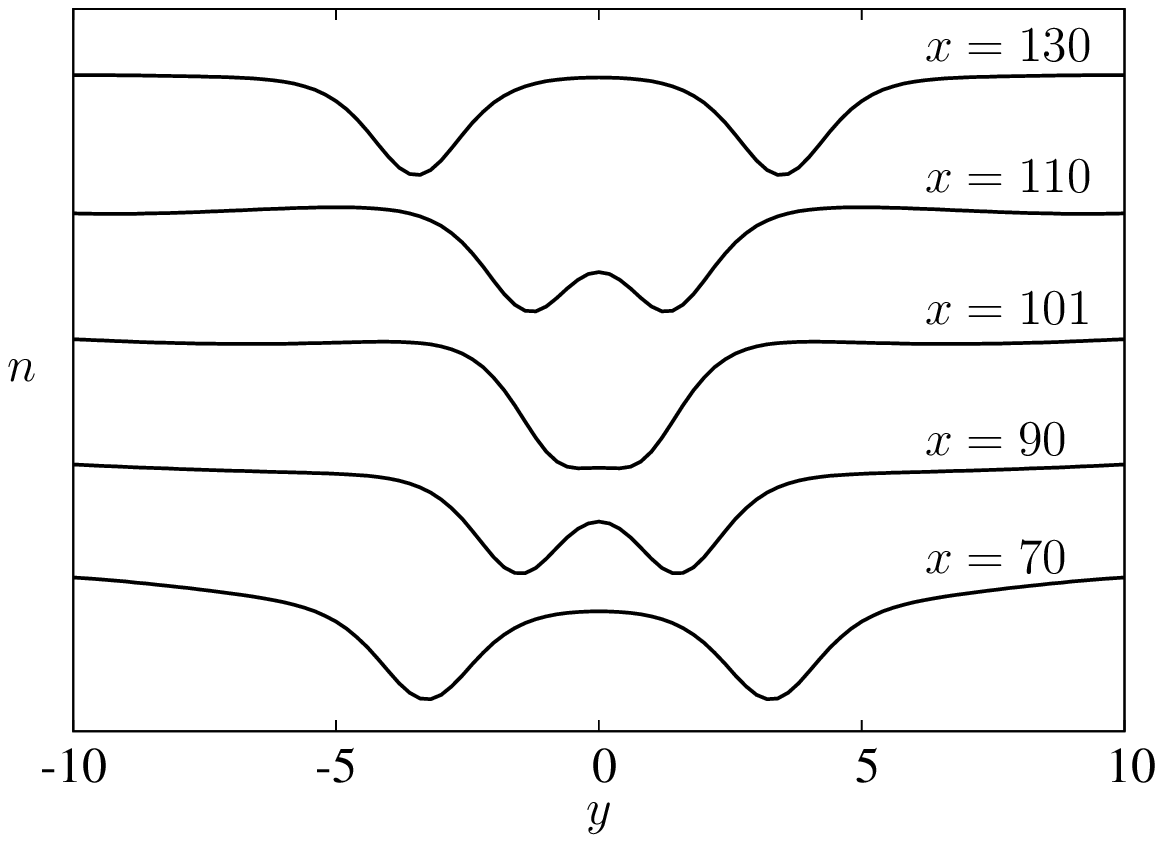}
\end{minipage} 
\hfill
\begin{minipage}[b]{0.45\linewidth}
\includegraphics[width=\linewidth]{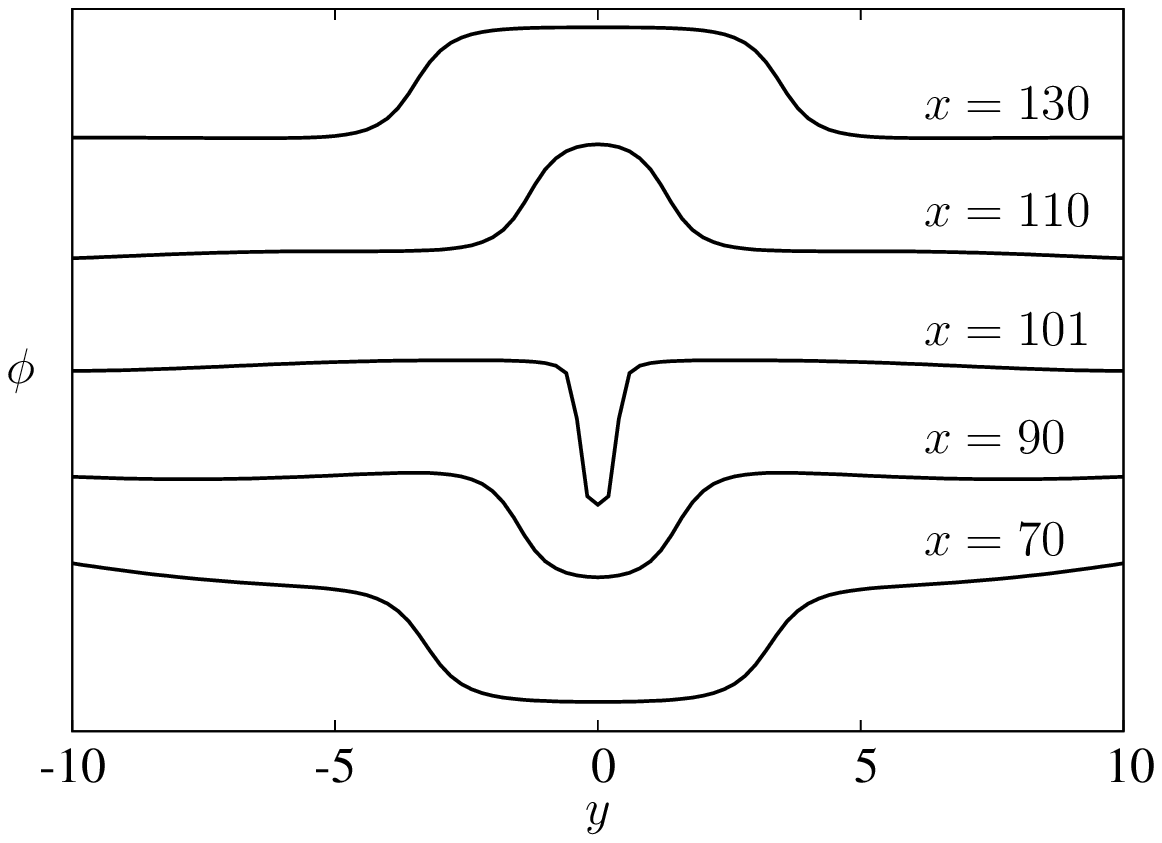}
\end{minipage} 
\caption{\label{dens_phase}
{\bf Left}: Cross sections of the density distribution shown in Fig. \ref{fig1}
for different values of $x$
obtained from the numerical solution of the 2D NLS equation (\ref{1-1}).
The value of $x$ is indicated on each curve.
The collision of two oblique dark solitons occurs at $x \approx 101$ and is practically
elastic.
{\bf Right}: Cross sections of the correspondent phase distribution. One can see a phase jump
after the collision of the two solitons
.}
\end{figure}
6. Now, we compare our analytical predictions with numerical simulations.
For large obstacles, many pairs of solitons can be generated at different angles past each 
obstacle \cite{egk-06}. For sake of simplicity, we consider here only small obstacles with 
radius $r\sim 1$, thus each obstacle only generates one pair of oblique dark solitons 
with angles $\theta$ and $-\theta$.
We present in Fig. \ref{dens_phase} (left) cross sections of the density distribution shown
in Fig. \ref{fig1} and the correpondent cross sections of the phase (right). 
The collision occurs at $x \approx 101$ and is practically elastic, i.e., we do not see any 
radiation loss during and after the solitons interaction. 
We also observe a phase jump after the collision.

In order to measure the shifts of the solitons positions after their 
interaction, we have performed two series of numerical simulations. 
Firstly, we have simulated the 2D flow past one impenetrable obstacle 
of unitary radius placed at $(x_{10},y_{10})=(0,-15)$ and for 
different values of the Mach number $M$, namely $M = 5,\, 6,\, 7,\, 8 
\,\,\rm{and}\,\, 10$. Then, we have measured the coordinates 
$({x}_{\rm 1A},{y}_{\rm 1A})$ of the minimum of the soliton far 
enough from the region of interaction, where subscript ``A" denotes 
simulation with one single obstacle.

Secondly, we added another obstacle at the position 
$(x_{20},y_{20})=(0,15)$, repeated the simulation of the 2D 
superfluid flow and measured the 
coordinates $({x}_{\rm 1B},{y}_{\rm 1B})$ of the minimum of the dark 
soliton, where ``B'' denotes simulation with two obstacles. Thus, for 
${x}_{\rm 1A} = {x}_{\rm 1B}$, the shift is given by  
$\Delta y_1= {y}_{\rm 1B} -{y}_{\rm 1A}$. In all simulations, we have applied a 2D 
finite difference method (Crank-Nicolson method) combined with a 
split-step method 
and used the spatial grid sizes $\delta x = \delta y = 0.2$ and the 
time step $\delta t = 0.01$. 

In Table 1 we compare the numerical results with the analytical 
predictions for the shifts using different values of $M$. The agreement 
with the analytical results is satisfactory considering the perturbation 
of the linear waves \cite{shipwaves} on the dark solitons and the computational 
limitation for the numerical simulations.


\begin{table}[ht]
\caption{Supersonic flow past two obstacles.} 
\centering  
\begin{tabular}{|c|c|c|c|} 
\hline\hline                       
$M$ & $\theta$ & $\Delta y_{1}$ & $\Delta y_{1}$ Eq.(\ref{5-1}) \\ [0.5ex] 
\hline                  
5  & 0.1  & 0.5 & 0.8             \\ \hline
6  & 0.08 & 0.6 & 0.8             \\ \hline
7  & 0.07 & 0.6 & 0.8  \\ \hline
8  & 0.06 & 0.6 & 0.8  \\ \hline  
10 & 0.05 & 0.5 & 0.8  \\ \hline 
\end{tabular}
\label{table:nonlin} 
\end{table}

\begin{figure}[t!!]
\begin{center}
\includegraphics[width=7.0cm,height=5cm]{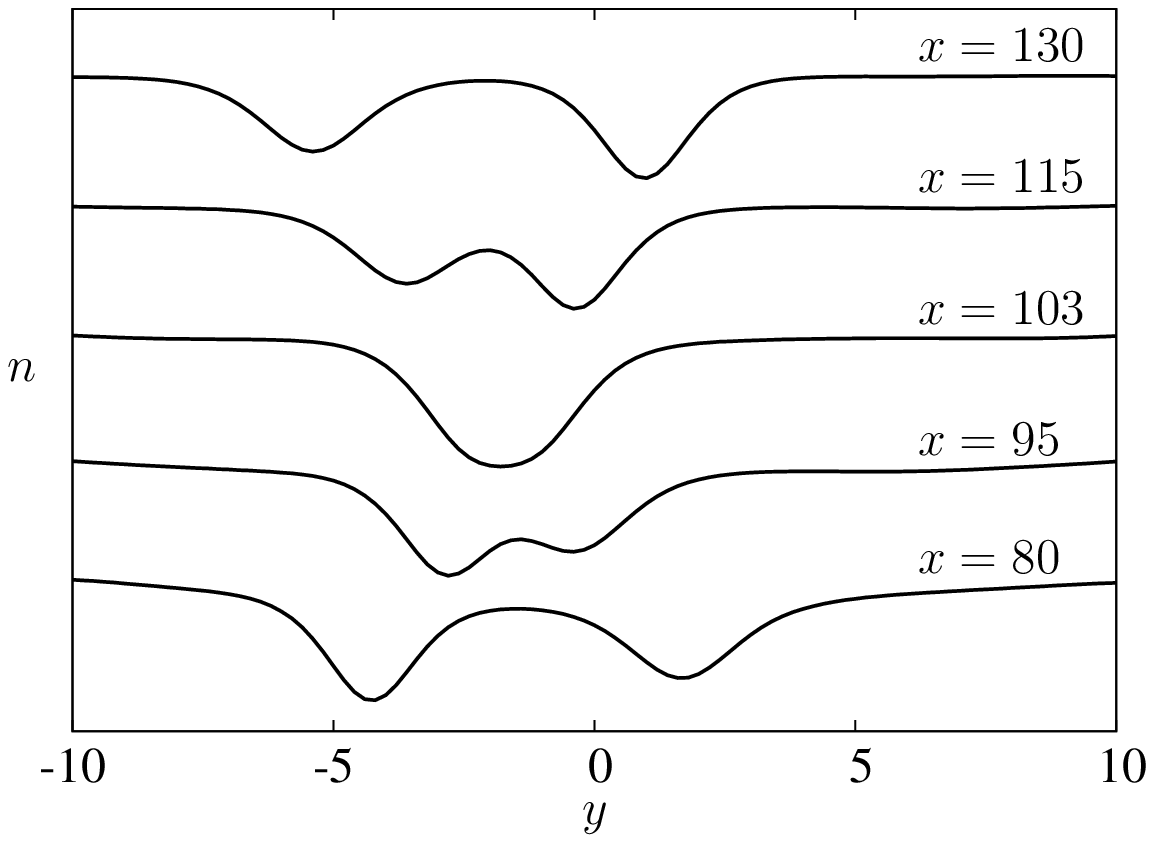}
\includegraphics[width=7.5cm,height=5.3cm]{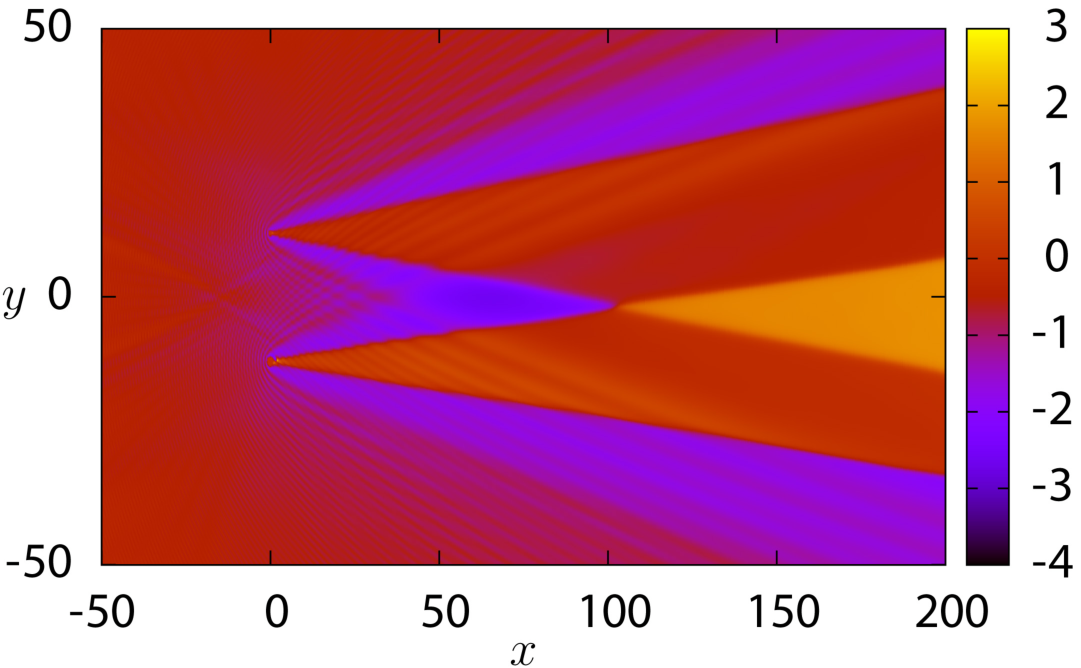}
\caption{\label{density2diff}
{\bf Left}: Cross sections of the density distribution for different values of $x$
obtained from the numerical solution of the 2D NLS equation (\ref{1-1}).
The value of $x$ is indicated on each curve.
We consider the flow past two different obstacles, one with radius $r=1.0$ at 
$(x_{10},y_{10})=(0,-12)$ and the other one with radius $r=0.6$ located at $(x_{20},y_{20})=(0,12)$.
The collision of two different oblique dark solitons occurs at $x \approx 103$ 
and is still practically elastic. The solitons angles are $\theta_1 = 0.1$ and
$\theta_2 = 0.13$ and the shifts of the solitons positions are 
$\Delta y_{1} \approx \Delta y_{2} \approx 0.5$.
{\bf Right}: Density plot of the corresponding phase. We can see a phase jump after the collision 
of the two solitons.
.}
\end{center}
\end{figure}

We further simulate the two-dimensional superfluid flow past a single small 
obstacle with different sizes and notice that the amplitudes and the slopes of 
these solitons depend on the obstacle's sizes. As it can be seen in Table 
\ref{T2do2}, increasing the size of the obstacle, the amplitude of the soliton 
increases and its slope decreases. Consequently, one can investigate the interaction of 
oblique dark solitons with different amplitudes and slopes considering two 
obstacles with different sizes. In Fig. \ref{density2diff} we show the 
superfluid flow past two obstacles, one with radius $r=0.6$ and the other one 
with radius $r=1$. We see that the collision of two 
different oblique dark solitons is still practically elastic and also the phase 
jump after the collision.

\begin{table}[ht]
\caption{Amplitude ($A$) and slope ($\theta$) of dark solitons for obstacles with different radius ($r$) and $M=5$.} 
\centering  
\begin{tabular}{|c|c|c|c|c|c|} 
\hline\hline                       
$r$     & $0.6$  & $0.8$  & $1.0$  & $1.2$  & $1.4$  \\ [0.5ex] 
\hline                  
$A$  & $0.57$ & $0.69$ & $0.78$ & $0.81$ & $0.83$ \\ \hline
$\theta$      & $0.13$ & $0.11$ & $0.1$  & $0.09$ & $0.08$  \\ \hline
\end{tabular}
\label{T2do2} 
\end{table}


\clearpage

7. Conclusions: We analyzed the collision of two oblique dark solitons 
numerically and by analytical approximations. 
The observed shifts are consistent in magnitude order with the analytical predictions,
considering the perturbation of the linear waves and the computational
limitation for the numerical simulations.
During and after the collision we have not observed any 
radiation loss and phase jumps are analogous to those observed in the 1D NLS. 
We conjecture that collisions of oblique solitons in 2D NLS may be a 
completely integrable process in the asymptotic limit. This soliton collision 
might be experimentally observed in different nonlinear media such as an atomic BEC, 
photorefractive crystals and exciton-polariton condensates.

We thank A.M Kamchatnov for useful discussions. We also thank funding agencies
CAPES, CNPq and FAPESP (Brazil).

\end{document}